\documentstyle[aps,epsfig,prl]{revtex}
\begin{document}
\title {The Kubo-type Formula for Conductivity of Spatially
Inhomogeneous Systems}
\author {I. G. Lang, L. I. Korovin}
\address {A. F. Ioffe Physical-Technical Institute, Russian
Academy of Sciences, 194021 St. Petersburg, Russia}
\author {S. T. Pavlov\dag\ddag}
\address {\dag Facultad de Fisica de la UAZ, Apartado Postal C-580,
98060 Zacatecas, Zac., Mexico \\
\ddag P.N. Lebedev Physical Institue of Russian Academy of
Sciences. 119991, Moscow, Russia}
\twocolumn[\hsize\textwidth\columnwidth\hsize\csname
@twocolumnfalse\endcsname
\date{\today}
\maketitle \widetext
\begin{abstract}
\begin{center}
\parbox {6in}
{The expressions for average densities of currents and charges
 induced by a weak electromagnetic field in
spatially inhomogeneous systems are obtained. The case of finite
temperatures is considered. It is shown that average values are
separated into "basic" and "additional" parts. The former depends
on electric fields, and the latter depends on derivatives of
electric fields on coordinates. Semiconductor quantum wells,
wires or dots may be considered as spatially inhomogeneous
systems.}
\end{center}
\end{abstract}
\pacs {PACS numbers: 78.47. + p, 78.66.-w}

] \narrowtext

1.R. J. Kubo [1] obtained  the formula for the conductivity tensor
$ \sigma _ {\alpha \beta} (\omega) $, applicable in a case of
spatially homogeneous systems and electric fields $ {\bf E} (t) $
independent of spatial coordinates. This formula takes into
account exactly the interaction of current carriers  with the
medium. Consequently, it is a powerful tool for the solution of
concrete conductivity problems in solids.

Currently there is a great interest in experimental and
theoretical study of low-dimensional semiconductor objects  such
as quantum wells, wires and dots. Because of this, the
generalization of the Kubo formula  in a case of spatially
inhomogeneous systems and spatially inhomogeneous  fields is
urgent.  Kubo [1] used the operator of interaction of current
carriers  with the  electric field as
\begin{equation}
\label{1} U _ K = -\sum _ ie _ i {\bf r} _ i {\bf E} (t),
\end{equation}
where $ e_i $ and $ {\bf r}_i $ are the charge and radius -
vector of the $ i $ -th particle, $ {\bf E}(t) $ is
time-dependent, but homogeneous in the space electric field.
However, it follows from the Maxwell equations, that the
time-dependent
 electric field necessarily depends
on coordinates, so the use of Eq. (1) is always a certain
approximation, if $ {\bf E} $ depends on $ t $. Our task consists
of obtaining the additional terms in the Kubo formula, which
contain derivatives of the electric field on coordinates, as well
as  taking into account the heterogeneity of physical systems.

2.Let us consider a system of $ N $ particles of charge $ e $ and
mass $ m $ in a weak electromagnetic field, characterized by
intensities
 $ {\bf E} ({\bf r}, t) $ and $ {\bf H} ({\bf r},
t) $. We introduce vector $ {\bf A} ({\bf r}, t) $ and scalar $
\varphi ({\bf R}, t) $ potentials in terms of which the fields
are expressed (the gage of potentials is arbitrary). For the sake
of generalizing our task, we shall suppose that the system of
particles is placed in a constant magnetic field $ {\bf H} $.
 The vector potential $ {\bf {\cal A}} ({\bf
r}) $ corresponds to ${\bf H}$. The total Hamiltonian $ {\cal
H}_{total} $ is as follows
\begin{equation}
\label{2} {\cal H}_{total} = {\cal H} + U,
\end{equation}
where
\begin{equation}
\label{3} {\cal H} = \sum_i {{\bf p}_i^2\over 2m} + V({\bf r}_
1\ldots {\bf r}_N),
\end{equation}
$ {\bf p} _ i = {\bf P} _ i - (e/c) {\bf {\cal A}} ({\bf r} _ i
), \, \, {\bf P} _ i = -i\hbar(\partial/\partial {{\bf r} _ i}),
$  $ V ({\bf r} _ 1\ldots {\bf r} _ N) $ is the potential energy,
including interaction between particles and external potential. $
U $ is the interaction energy of particles with the
electromagnetic field
\begin{equation}
\label{4} U = U _ 1 + U _ 2,
\end{equation}
\begin{equation}
\label{5} U_1 =-{1\over c}\int d^3r{\bf j}(r){\bf A}({\bf r}, t)
+ \int d^3r\rho ({\bf r})\varphi({\bf r}, t),
\end{equation}
\begin{equation}
\label{6} U _ 2 = {e\over 2mc} \int d^3r \, \rho (r) \, {\bf A} ^
2 ({\bf r}, t).
\end{equation}
In  Eqs. (5), (6) the operators of current density
$$ {\bf j} ({\bf r}) = \sum _ i {\bf j} _ {\, i \,} ({\bf r}), $$
$$ {\bf j} _ {\, i \,} ({\bf r}) = {e\over 2}\{\delta ({\bf r} - {\bf r} _ i) {\bf
v} _ i + {\bf v} _ i \delta ({\bf r} - {\bf r} _ i) \},~~~{\bf v}
_ i = {{\bf p} _ i\over m } $$  and charge density
$$\rho ({\bf r}) = \sum _ i\rho _ i ({\bf r}), \quad \rho _ i ({\bf r}) =
e\delta ({\bf r} - {\bf r} _ i) $$ are introduced. The operator $
U _ 2 $ is out of the  framework of the linear approximation on
the electromagnetic field.

It is supposed that on the indefinitely removed distances there
are no charges and currents, and also that on times $ t\to
-\infty, $ the fields $ {\bf E} ({\bf r}, t) $ and $ {\bf H} ({\bf
r}, t) $ are equal to $ 0 $, which corresponds to adiabatical
switching on the fields. Generally, the operator $ U $ of the
interaction of particles with the electromagnetic field is not
expressed through fields $ {\bf E} ({\bf r}, t) $ and $ {\bf H}
({\bf R}, t) $. Accordingly, the operators $ {\bf j} _ 1 ({\bf r},
t) $ and $ \rho _ 1 ({\bf r}, t) $ are expressed only through
potentials $ {\bf A} ({\bf r}, t) $ and $ \varphi ({\bf r}, t) $
as follows
\begin{eqnarray}
\label{7}{\bf j}_{1\alpha}({\bf r}, t)&=&-{e\over mc}\rho({\bf r},
t)A_\alpha({\bf r}, t)+{i\over\hbar}\int_{-\infty}^tdt^\prime
[U_1(t^\prime), j_\alpha({\bf r}, t)],\nonumber\\
\rho_1({\bf r}, t)&=&{i\over\hbar}\int_{-\infty}^t
dt^\prime[U_1(t^\prime), \rho({\bf r}, t)],
\end{eqnarray}
where subscript $ 1 $ means linear approximation on the fields. $
[F, Q] = FQ-QF $ is the commutator of operators $ F $ and $ Q $.

However, all the observable values must be expressed through
fields. This also is true for average values $ \langle j _
{1\alpha} ({\bf r}, t) \rangle $ and $ \langle \rho _ 1 ({\bf r},
t) \rangle $, where the averaging is introduced by
\begin{equation}
\label{8} <\ldots> = \frac {Sp \{\exp (-\beta {\cal H})\ldots \}}
{Sp \{\exp (-\beta {\cal H}) \}}, \quad \beta = \frac {1} {kT}.
\end{equation}

The problem lies in what kind of interaction to use a containing
vector potential or an electric field, which was discussed earlier
in [2] with reference to a task of the light  dispersion in bulk
crystals. However, since we must solve other problems with
spatially inhomogeneous systems, it is necessary to come back
again  to this topic.

3.For the case $T=0$ average values of densities of current and
charge, induced by external fields in a spatially inhomogeneous
medium,  are calculated in [3]. The obtained expressions contain
the symbol $ \langle 0 |... | 0\rangle \, $, where $ | 0\rangle \,
$ is the wave function of the ground state of the system. By
replacing averaging $ \langle 0 |... | 0\rangle \, $ with
averaging Eq. (8) [3] we obtain
\begin{equation}
\label{9} \langle \, j _ {1\alpha} ({\bf r}, t) \rangle \,
=\langle \, j _ {1\alpha} ({\bf r}, t) \rangle \, _ E + \langle
\, j _ {1\alpha} ({\bf r}, t) \rangle \, _ {\partial E /\partial
r},
\end{equation}
\begin{equation}
\label{10} \langle \,\rho _ 1 ({\bf r}, t) \rangle \, =\langle
\,\rho _ 1 ({\bf r}, t) \rangle \, _ E + \langle \,\rho _ 1 ({\bf
r}, t) \rangle \, _ {\partial E /\partial r},
\end{equation}
where
\begin{eqnarray}
\label{11} \langle\,j_{1\alpha}({\bf
r},t)\rangle\,_E={i\over\hbar}\int\,d^3r^\prime
\int_{-\infty}^t\,dt^\prime\nonumber\\
\times \langle\,[j_\alpha({\bf r},t),\,d_\beta({\bf
r}^\prime,t^\prime)]\rangle\, E_\beta({\bf r}^\prime,t^\prime),
\end{eqnarray}
\begin{eqnarray}
\label{12} \langle \,\rho _ 1 ({\bf r}, t) \rangle \, _ E =
{i\over\hbar}\int \, d ^ 3r ^ \prime
\int _ {-\infty} ^ t \, dt ^ \prime\nonumber\\
\times \langle \, [\rho ({\bf r}, t), \, d _ \beta ({\bf r} ^
\prime, t ^ \prime)] \rangle \, E _ \beta ({\bf r} ^ \prime, t ^
\prime),
\end{eqnarray}
\begin{eqnarray}
\label{13} \langle \, j _ {1\alpha} ({\bf r}, t) \rangle \, _
{\partial E /\partial r} = {e\over mc} \langle \, d _ \beta ({\bf
r}) \rangle \, \frac {\partial a _ \beta ({\bf r}, t)} {\partial
r _ \alpha} \nonumber\\-
 {i\over\hbar c}\int \, d ^ 3r ^ \prime\int _ {-\infty} ^ t \, dt ^ \prime
\langle \, [j _ \alpha ({\bf r}, t), Y _ {\beta\gamma} ({\bf r} ^
\prime, t ^ \prime)] \rangle \nonumber\\
\times\frac {\partial a _ \beta ({\bf r} ^ \prime, t ^ \prime)}
{\partial r _ \gamma ^ \prime},
\end{eqnarray}
\begin{eqnarray}
\label{14} \langle \,\rho _ 1 ({\bf r}, t) \rangle \, _ {\partial
E /\partial r} = -{i\over\hbar c}\int \, d ^ 3r ^ \prime\int _ {-\infty} ^ t \, dt ^ \prime\nonumber\\
\times\langle \, [\rho ({\bf r}, t), Y _ {\beta\alpha} ({\bf r} ^
\prime, t ^ \prime)] \rangle \, \frac {\partial a _ \beta ({\bf r}
^ \prime, t ^ \prime)} {\partial r _ \gamma ^ \prime},
\end{eqnarray}
\begin{equation}
\label{15} {\bf d }({\bf r}) = {\bf r} \, \rho ({\bf r}), \qquad Y
_ {\beta\gamma} ({\bf r}) = r _ \beta \, j _ \gamma ({\bf r}),
\end{equation}
\begin{equation}
\label{16} {\bf a} ({\bf r}, t) = -c\int _ {-\infty} ^ t dt ^
\prime {\bf E} ({\bf r}, t ^ \prime) .
\end{equation}

The correctness of the replacement of averaging $ \langle \, 0
|... | 0\rangle \, $ with $ \langle \, ...\rangle \, $ proves to
be true in the further comparison of our results with the results
of [4] and [5].

4. We transform Eqs. (9) -- (14) so that at transition to a limit
$ {\bf E} ({\bf r}, t) \simeq {\bf E} (t), $  the Kubo formula
 for conductivity [1] will be obtained. We use a ratio [1,4]
\begin{equation}
\label{17} {i\over\hbar}\langle[F (t), Q (t^\prime)]\rangle=
\int_0^\beta d\lambda\langle{dQ(t^\prime)\over dt^\prime}F(t+
i\hbar\lambda)\rangle ,
\end{equation}
true for any pair of operators $ F $ and $ Q $, and we also use
the integration on variable $ t ^ \prime $ and $ {\bf r} ^ \prime
$ in parts. It results in
\begin{equation}
\label{18} \langle \, j _ {1\alpha} ({\bf r}, t) \rangle \,
=\langle \, j _ {1\alpha} ({\bf r}, t) \rangle \, ^ {(1)} +
\langle \, j _ {1\alpha} ({\bf r}, t) \rangle \, ^ {(2)},
\end{equation}
where the first part contains an electric field, and the second
contains its derivative on coordinates, i.e.
\begin{eqnarray}
\label{19} \langle \, j _ {1\alpha} ({\bf r}, t) \rangle \, ^
{(1)} = \int \, d ^ 3r ^ \prime
\int _ {-\infty} ^ t \, dt ^ \prime\int _ 0 ^ \beta \, d\lambda \,\nonumber\\
\times\langle \, j _ \beta ({\bf r} ^ \prime, t ^ \prime) \, j _
\alpha ({\bf r}, t + i\hbar\lambda) \rangle \, E _ \beta ({\bf r}
^ \prime, t ^ \prime),
\end{eqnarray}
\begin{eqnarray}
\label{20} \langle \, j _ {1\alpha} ({\bf r}, t) \rangle^
{(2)}={e\over mc}\langle d_\beta({\bf r})\rangle\frac{\partial
a_\beta({\bf r}, t)}{\partial r_alpha}\nonumber
\\-
 {1\over c}\int d^3r^\prime\int_0^\beta d\lambda
\langle Y_{\beta\gamma}({\bf r}^\prime)j_\alpha({\bf
r}, i\hbar\lambda)\rangle\nonumber\\
\times\frac{\partial a_\beta({\bf r}^\prime, t)}{\partial r_
\gamma^\prime}.
\end{eqnarray}
Similarly, we obtain
\begin{equation}
\label{21} \langle \,\rho _ 1 ({\bf r}, t) \rangle \, =\langle
\,\rho _ 1 ({\bf r}, t) \rangle \, ^ {(1)} + \langle \,\rho _ 1
({\bf r}, t) \rangle \, ^ {(2)},
\end{equation}
\begin{eqnarray}
\label{22} \langle \,\rho _ 1 ({\bf r}, t) \rangle \, ^ {(1)} =
\int \, d ^ 3r ^ \prime
\int _ {-\infty} ^ t \, dt ^ \prime\int _ 0 ^ \beta \, d\lambda \,\nonumber\\
\times \langle \, j _ \beta ({\bf r} ^ \prime, t ^ \prime) \,
\rho ({\bf r}, t + i\hbar\lambda) \rangle \, E _ \beta ({\bf r} ^
\prime, t ^ \prime),
\end{eqnarray}
\begin{eqnarray}
\label{23} \langle \,\rho _ 1 ({\bf r}, t) \rangle \, ^ {(2)} =
-{1\over c}\int \, d ^ 3r ^ \prime\int _ 0 ^ \beta \, d\lambda \, \nonumber\\
\times\langle \, Y _ {\beta\gamma} ({\bf r} ^ \prime) \, \rho
({\bf r}, i\hbar\lambda) \rangle \, \frac {\partial a _ \beta
({\bf r} ^ \prime, t)} {\partial r _ \gamma ^ \prime}.
\end{eqnarray}
Let us emphasize that the separations of the average values Eqs.
(18) and (21)  in two parts does not coincide with the separations
Eqs. (9) and (10), convenient only at $ T = 0 $.

The contributions with the superscript (1) contain an electric
field. We shall name them "basic". The contributions with the
superscript (2) contain derivatives of the electric field on
coordinates.  We shall name them "additional". The reason for
introducing  basic and additional contributions is that for the
solution of any problem, in which the field $ {\bf E} ({\bf r}, t)
$ is spatially inhomogeneous, it is possible to estimate
magnitudes of the additional contributions and to determine
whether it is necessary  to take them into account or to reject
them.

5.We compare our results with the conclusions of [5], where $
\langle j _ {1\alpha} ({\bf r}, t) \rangle $ is also calculated at
finite temperatures.   This expression differs from ours which
only  contains an electric field. Generally speaking, integrating
on $ {\bf r} ^ \prime $ in parts, it is possible to get rid of
derivative $
\partial E _ \beta ({\bf r} ^ \prime, t) /\partial r _ \gamma ^
\prime $, proceeding to the formulas which only  contain fields.
However, the opposite procedure, transition from fields to
derivatives, is  not always possible in basic contributions where
fields are always kept.

The separation of Eq. (18) in two parts coincides with the
separation of [5]. The expressions for the basic contribution
with superscript (1) coincide. But instead of Eq. (20) in [5] for
the additional contribution, the result
\begin{eqnarray}
\label{24} {\partial \langle \, j _ {1\alpha} ({\bf r}, t)
\rangle ^ {(2)} \over\partial t} = -\int d^3r^\prime \,\varphi _
{\alpha\beta} ({\bf r}, {\bf
r} ^ \prime, 0) \nonumber\\
\times [E _ \beta ({\bf r} ^ \prime, t) -E _ \beta ({\bf r}, t,)]
\end{eqnarray}
is obtained, where
\begin{equation}
\label{25} \varphi _ {\alpha\beta} ({\bf r}, {\bf r} ^ \prime,
\tau) = \int _ 0 ^ \beta \, d\lambda \,\langle \, j _ \beta ({\bf
r} ^ \prime) \, j _ \alpha ({\bf r}, \tau + i\hbar\lambda)
\rangle.
\end{equation}
Integrating both parts of Eq. (24) on time, we have
\begin{eqnarray}
\label{26} \langle \, j _ {1\alpha} ({\bf r}, t) \rangle \, ^
{(2)} = - {e\over mc}\langle \,\rho ({\bf r}) \rangle \, a _
\alpha ({\bf r}, t) \nonumber\\ +
 {1\over c}\int \, d ^ 3r ^ \prime\int _ 0 ^ \beta \, d\lambda
\langle \, j _ \beta ({\bf r} ^ \prime) j _ \alpha ({\bf r},
i\hbar\lambda) \rangle \, a _ \beta ({\bf r} ^ \prime, t).
\end{eqnarray}
Similarly, we obtain
\begin{eqnarray}
\label{27} \langle \,\rho ({\bf r}, t) \rangle \, ^ {(2)} =
{1\over c} \int \, d ^ 3r ^ \prime\int _ 0 ^ \beta \, d\lambda \,\nonumber\\
\langle \, j _ \beta ({\bf r} ^ \prime) \, \rho ({\bf r},
i\hbar\lambda) \rangle \, a _ \beta ({\bf r} ^ \prime, t).
\end{eqnarray}
We have verified that  it is possible to proceed from our results
in Eqs. (20) and (23) to the results of Eqs. (26) and (27),
integrating on $ {\bf r} ^ \prime $ in parts.

6. Eqs. (20) and (23), unlike Eqs. (26) and (27), contain
operators $ {\bf r} _ i $ of coordinates of particles. Indeed, the
definitions in Eqs. (15) may be written as
\begin{equation}
\label{28} {\bf d} ({\bf r}) = e \,\sum _ i \, {\bf r} _ i
\,\delta ({\bf r} - {\bf r} _ i),
\end{equation}
\begin{equation}
\label{29} Y_{\beta\gamma}({\bf r})={e\over 2}\sum_i[r_
{i\beta}j_{i\gamma}+j_{i\gamma} r_{i \beta}].
\end{equation}
But the average values $ \langle \, {\bf j} _ 1 ({\bf r}, t)
\rangle \, $ and $\langle \rho _ 1 ({\bf r}, t)\rangle $ should
not depend on the point of readout of the coordinates $ {\bf r} _
i $. This means that Eqs. (20) and (23) contain only non-diagonal
elements of the operators $ {\bf r} _ i $, and the diagonal
elements can be excluded.

To demonstrate this statement let us transform Eq. (29) so that
the operator $ r _ {i\beta} $ stands only on the left
\begin{equation}
\label{30} Y _ {\beta\gamma} ({\bf r}) = - {i\hbar\over 2m} \,
\delta _ {\beta\gamma} \, \rho ({\bf r}) \, + \, \sum _ i \, r _
{i\beta} \, j _ {i\gamma} ({\bf r}).
\end{equation}
Substituting Eqs. (28) and (30) in Eq. (20), we obtain
\begin{eqnarray}
\label{31} \langle \, j _ {1\alpha} ({\bf r}, t) \rangle \, ^
{(2)} = { e^2\over mc} \sum _ i \,\langle \, r _ {i\beta} \,
\delta ({\bf r} - {\bf r} ^ \prime) \rangle \, \frac {\partial a _
\beta ({\bf r}, t)} {\partial r _ \alpha} \nonumber\\ +
{i\hbar\over 2mc} \int \, d ^ 3r ^ \prime \,\int _ 0 ^ \beta \, d
\,\lambda \, \langle \,\rho ({\bf r} ^ \prime) \, j _ \alpha \,
({\bf r}, i\hbar\lambda) \rangle \nonumber\\
\times div \, {\bf a} ({\bf r} ^ \prime, t) \nonumber\\- {1\over
c} \int \, d ^ 3r ^ \prime \,\int _ 0 ^ \beta \, d \,\lambda
\,\sum _ i \, \langle \, r _ {i\beta} \, j _ {i\gamma} ({\bf r} ^
\prime) \, j _ \alpha ({\bf r}, i\hbar\lambda) \rangle \nonumber\\
\times\frac {\partial a _ \beta ({\bf r} ^ \prime, t)} {\partial r
^ \prime _ \gamma}.
\end{eqnarray}
Let us separate the operator $ {\bf r} _ i $ in two parts:
\begin{equation}
\label{32} {\bf r} _ i \, = \, {\bf r} _ i ^ d \, + \, {\bf r} _ i
^ {nd},
\end{equation}
where superscripts $ d $ and $ nd $ mean diagonal and
non-diagonal contributions, respectively, and
\begin{equation}
\label{33} \langle n |{\bf r}_i^d | m\rangle = \langle n | {\bf
r}_i | m\rangle \langle  n | m\rangle.
\end{equation}
We then substitute Eq. (32) in the last term  of Eq. (31) from
item 4. Taking advantage of commutativity  $ [{\cal H}, {\bf r} _
i ^ d] \, = \, 0, $ it is possible to show that
\begin{eqnarray}
\label{34} -{1\over c}\int d^3r^\prime\int_0^\beta
d\lambda\sum_i\langle r_{i\beta}^d j_{i\gamma}({\bf r}^\prime)
j_\alpha({\bf r},i\hbar\lambda)\rangle\nonumber\\
\times \frac{\partial a_\beta({\bf r}^\prime,t)}{\partial
r^\prime_\gamma}= -{e\over mc}\sum_i\langle r_{i\beta}\delta({\bf
r}-{\bf r}_i)\rangle{\partial a_\beta({\bf r},t)\over\partial
r_\alpha}\nonumber\\+ {1\over c}\int d^3r^\prime \int_0^\beta
d\lambda\sum_i\langle j_{i\gamma}({\bf r}^\prime)j_\alpha({\bf
r},i\hbar\lambda)
r_{i\beta}^{nd}(i\hbar\lambda)\rangle\nonumber\\
\times\frac{\partial a_\beta({\bf r}^\prime,t)}{\partial
r^\prime_\gamma}.
\end{eqnarray}
Using Eqs. (32) and (34), we obtain from Eq. (31) expressions,
which do not containing the operators $ {\bf r} _ i ^ d, $
\begin{eqnarray}
\label{35} \langle\,j_{1\alpha}({\bf r},t)\rangle\,^{(2)}\,=\,
{i\hbar\over 2mc}\int\,d^3r^\prime\,\int_0^\beta\,d\,\lambda\,\nonumber\\
\times\langle\,\rho({\bf r}^\prime)\, j_\alpha({\bf
r},i\hbar\lambda)\rangle\,\,div\,{\bf a}({\bf
r}\prime,t)\nonumber\\+ {1\over c}\int d^3r^\prime\int_0^\beta
d\lambda\sum_i\langle j_{i\gamma}({\bf r}^\prime)j_\alpha({\bf
r}, i\hbar\lambda)\,
r_{i\beta}^{nd}(i\hbar\lambda)\rangle\nonumber\\-
 r_{i\beta}^{nd} j_{i\gamma}^\prime({\bf
r}^\prime)j_\alpha({\bf r},i\hbar\lambda)\frac{\partial
a_\beta({\bf r}^\prime,t)}{\partial r^\prime_\gamma}.
\end{eqnarray}
Similarly, for the additional contribution in average induced
density of charge we have
\begin{eqnarray}
\label{36} \langle \,\rho ({\bf r}, t) \rangle \, ^ {(2)} = \frac
{i\hbar} {2mc} \int \, d ^ 3r ^ \prime
\int _ 0 ^ \beta \, d\lambda\nonumber\\
\times\langle \,\rho ({\bf r} ^ \prime) \, \rho ({\bf r}, i\hbar
\lambda) \rangle \, div \, {\bf a} ({\bf r} ^ \prime, t)
\nonumber\\ + \frac {1} {c} \int \, d ^ 3r ^ \prime\int _ 0 ^
\beta \, d\lambda\sum _ i \langle \, j _ {i\gamma} ({\bf r} ^
\prime) \rho ({\bf r}, i\hbar \lambda) \, r _ {i\beta} ^ {nd}
(i\hbar\lambda) \nonumber\\- r _ {i\beta} ^ {nd} j _ {i\gamma}
({\bf r} ^ \prime) \, \rho ({\bf r}, i\hbar\lambda) \rangle \,
\frac {\partial a _ \beta ({\bf r} ^ \prime, t)} {\partial r _
\gamma ^ \prime}.
\end{eqnarray}

7. We make the Fourier-transformation of the electric field as
follows
\begin{eqnarray}
\label{37} E_\alpha({\bf r},t)&=&E_\alpha^{(+)}({\bf r},t)\,+\,
E_\alpha^{(-)}({\bf r},t),\nonumber\\
E_\alpha^{(+)}({\bf r},t)&=&{1\over (2\pi)^{4}}\int
d^3k\int_0^\infty\,d\omega E_\alpha({\bf k},\omega)e^{i({\bf
k}{\bf r}-\omega t)},\nonumber\\
E_\alpha^{(-)}({\bf r},t)&=&(E_\alpha^{(+)}({\bf
r},t))^*,\nonumber\\
 E_\alpha({\bf
k},\omega)&=&\int\,d^3r\,\int_{-\infty}^\infty\,dt\,
E_\alpha({\bf r},t)\,e^{-i({\bf k}{\bf r}-\omega t)}.
\end{eqnarray}
The average induced density of current may now be written as
\begin{eqnarray}
\label{38} \langle \, j _ {1\alpha} ({\bf
r}, t) \rangle \, \, = \, {1\over (2\pi)^{4}} \int \, d ^ 3k \,\int _ 0 ^ \infty \, d\omega\nonumber\\
\times\sigma _ {\alpha\beta} ({\bf k}, \omega \, | \, {\bf r}) \,
E _ \beta ({\bf k}, \omega) \, e ^ {i ({\bf k} {\bf r} -\omega
t)} \, + \, c.c.,
\end{eqnarray}
where $ \sigma _ {\alpha\beta} ({\bf k}, \omega \, | \, {\bf r})
$ is the conductivity tensor, dependent on spatial coordinates.
 Using Eqs. (19) and (35) for
basic and additional contributions to the average induced density
of current, we obtain
\begin{equation}
\label{39} \sigma _ {\alpha\beta} ({\bf k}, \omega \, | \, {\bf
r}) = \sigma _ {\alpha\beta} ^ {(1)} ({\bf k}, \omega \, | \,
{\bf r}) + \sigma _ {\alpha\beta} ^ {(2)} ({\bf k}, \omega \, |
\, {\bf r}),
\end{equation}
where
\begin{eqnarray}
\label{40} \sigma _ {\alpha\beta} ^ {(1)}\, {\bf k}, \omega \, |
\, {\bf r}) = \int \, d ^ 3r ^ \prime \int _ 0 ^ \infty \, dt\int
_ 0 ^ \beta
d\lambda\nonumber\\
\times \langle \, j _ \beta ({\bf r} - {\bf r} ^ \prime,
-i\hbar\lambda) \, j _ \alpha ({\bf r}, t) \rangle \, e ^ {-i
({\bf k} {\bf r} ^ \prime-i\omega t)},
\end{eqnarray}
\begin{eqnarray}
\label{41} \sigma _ {\alpha\beta} ^ {(2)}\, {\bf k}, \omega \, |
\, {\bf r}) = \frac {i\hbar k _ \beta} {2m\omega} \int \, d ^ 3r ^
\prime \, e ^ {-i {\bf
k} {\bf r} ^ \prime} \int _ 0 ^ \beta \, d\lambda\nonumber\\
\times \langle \,\rho ({\bf r} - {\bf r} ^ \prime, -
i\hbar\lambda) j _ \alpha ({\bf r}) \rangle\nonumber\\ +
 \frac {k _ \gamma} {\omega} \int \, d ^ 3r ^ \prime \, e ^ {-i {\bf
k} {\bf r} ^ \prime} \int _ 0 ^ \beta \, d\lambda\sum_i \langle \,
j _ {i\gamma} ({\bf r} - {\bf r} ^ \prime, -i\hbar\lambda) j _
\alpha ({\bf r}) r _ {i\beta} ^ {nd} \nonumber\\-r _ {i\beta} ^
{nd} (-i\hbar\lambda) j _ {i\gamma} ({\bf r} - {\bf r} ^ \prime,
-i\hbar\lambda) j _ \alpha ({\bf r}) \rangle.
\end{eqnarray}

As an approximation, when the electric field does not depend on
spatial coordinates, i.e. $ {\bf E} ({\bf r}, t) \simeq {\bf E}
(t) $, we make the Fourier-transformation
\begin{eqnarray}
\label{42} E _ \alpha ^ {(+)} (t) \, = {1\over 2\pi}\int _ 0 ^
\infty \, d\omega \,
E _ \alpha \, (\omega) \, e ^ {-i\omega \, t}, \nonumber\\
 \langle \, j _ {1\alpha} ({\bf r}, t) \rangle \, _ h \, = \,
{1\over 2\pi}\int _ 0 ^ \infty \, d\omega\nonumber\\
\times\sigma _ {\alpha\beta} (\omega \, | \, {\bf r}) \, E _
\beta \, (\omega) \, e ^ {-i\omega \, t} \, + \, c.c.,
\end{eqnarray}
where subscript $ h $ means a field, homogeneous in space. It is
easy to see that
\begin{equation}
\label{43} \sigma _ {\alpha\beta} (\omega \, | \, {\bf r}) =
\sigma _ {\alpha\beta} \, ({\bf k} = 0, \omega \, | \, {\bf r}).
\end{equation}
Then with the help of Eqs. (40) - (42) we obtain
\begin{equation}
\label{44} \sigma_{\alpha\beta}(\omega |{\bf r})=\int_0^\infty
dt\int _ 0 ^ \beta d\lambda\langle J _ \beta (-i\hbar \lambda) j
_ \alpha ({\bf r}, t) \rangle e ^ {i\omega t},
\end{equation}
where the current operator is
\begin{equation}
\label{45} J _ \alpha = e\sum _ i \,\dot {{\bf r}} _ {i\alpha}.
\end{equation}
Eq. (44) is the generalization of the Kubo formula  in the case of
a spatially inhomogeneous medium, where the conductivity tensor
depends on $ {\bf r}. $ In the case of a spatially homogeneous
medium, the tensor $ \sigma _ {\alpha\beta} (\omega | {\bf r}) $
does not depend on $ {\bf r} $ and we obtain from Eq. (44)
\begin{equation}
\label{46} \sigma _ {\alpha\beta}(\omega) = \frac {1} {V _ 0}
\int _ 0 ^ \infty dt \int _ 0 ^ \beta \, d\lambda\langle J _
\beta (-i\hbar\lambda) J _ \alpha (t) \rangle e ^ {i\omega t},
\end{equation}
where $ V _ 0 $ is the normalization volume. The obtained formula
coincides with the Kubo result [1] if we substitute   $ \omega $
for $ -\omega $ on the RHS and take into account that in [1] the
volume $ V _ 0 = 1 $.

It follows from Eq. (42) that the additional contributions to the
average density of the induced current result in additional
contributions to the conductivity tensor which contains the factor
$ k _ \gamma/\omega $. If a field $ {\bf E} ({\bf r}, t) $ is the
plane wave (at a monochromatic irradiation) or wave package ( at a
pulse irradiation), $ k\simeq\omega/c $ and additional
contributions  contain a small factor $ v/c $ in comparison to the
basic contributions, where $ v $ is the particle velocity in the
system.

However, this estimation is not always correct   for spatially
inhomogeneous systems (for example for semiconductor quantum
wells, wires or dots). It is possible to consider the field $
{\bf E} ({\bf r}, t) $ as external, or as a stimulating field,
only  when we calculate the densities of the induced current and
charge in the lowest order on interaction of the field with the
system of charged particles. Such an approximation is acceptable
 in the case of a quantum well under the condition [6,7]
$$\gamma _ r \,\ll \,\gamma, $$
where $ \gamma _ r (\gamma) $ is the radiative (non-radiative)
broadening of electronic excitation.

Otherwise, when $ \gamma _ r\gg\gamma, $ it is necessary to take
into account the interaction of the  field with particles in all
orders of a perturbation theory. Then the field $ {\bf E} ({\bf
r}, t) $ is the genuine field within the limits of
low-dimensional objects.

 This field cannot yet
be represented as  a superposition of plane waves, for which $ k
= \omega /c $. For example, the genuine field strongly varies
inside of a quantum well along the  $ z $ axis which is
perpendicular to the plane of a quantum well. The light falls
along the $ z $ axis and the frequency $\omega$ is in resonance
with one of the excitation energy levels of the electron system in
a quantum well belonging to the discrete spectrum [8,9]. Then the
values $ kd\simeq 1 $ are essential, where $ d $ is the width of
a quantum well. Instead of a small factor $ v/c, $  we obtain a
factor
$$M\simeq\frac{v}{\omega d}=\frac{v}{c}\frac{\lambda}{2\pi d}.$$

If the wave  length is $ \lambda\gg d $, it may appear that the
new factor $ M $ is much greater than $ v/c $. In concrete cases,
it is necessary to estimate its magnitude. In [8,9] we assumed
that $ M\ll 1 $ and neglected the additional contributions to
average values of the induced densities of currents and charges
(it was considered in the case T = 0).

Thus, Eqs. (40) - (42) represent the analogue of the Kubo formula
in a case of spatially inhomogeneous electric fields and systems.
The average values of the induced charge density are also
calculated in this case (see Eqs. (21), (22) and (37)).

The problem of the transition from initial expressions (containing
vector and scalar potentials)  to expressions for average values
(containing electric fields and derivatives from field on
coordinates) is solved.

Substituting the obtained expressions for the average densities of
current and charge in the Maxwell equations, it is possible in
principle to determine the genuine fields inside and outside of
semiconductor low-dimensional objects. Thus, it is possible to
calculate the light reflection and absorption coefficients  of
these objects (see, for example, [8,9]).

      Authors are grateful to M. Harkins for a critical reading of the
manuscript. The work has received partial financial support of
the Russian Fund of Basic Researches (00-02-16904), Program  "
Physics of semiconductor nanostructures" and Federal Program
"Integration".

\end{document}